\documentclass[reprint, prd]{revtex4-2}
\usepackage{amsmath, amssymb}
\usepackage{graphicx,color,float,tabularx,siunitx}
\usepackage[caption=false]{subfig}
\usepackage{placeins}
\usepackage{rviewport}
\usepackage{orcidlink}
\definecolor{colorLink}{rgb}{0,0,180} 
\usepackage{hyperref}
\hypersetup{
   colorlinks = true,
   citecolor  = colorLink,
   urlcolor   = colorLink,
   linkcolor  = colorLink,
}

\DeclareSIUnit\year{yr}

\usepackage{float} 
\usepackage{graphicx}
\usepackage{subcaption}

\usepackage[sort&compress]{natbib}

\usepackage[shortcuts]{extdash}

\begin{document}
\title{Degenerate Bifurcations and Universal Relaxation Scaling in Black Hole Thermodynamics
}

\author{Bidyut Hazarika \orcidlink{0009-0007-8817-1945}$^1$}

\email{bidyuthazarika1729@gmail.com}
\author{Mozib Bin Awal \orcidlink{0009-0007-8817-1945}$^1$}

\email{rs\_mozibbinawal@dibru.ac.in}

\author{Prabwal Phukon \orcidlink{0000-0002-4465-7974}$^1$,$^2$}
\email{prabwal@dibru.ac.in}

	\affiliation{$^1$Department of Physics, Dibrugarh University, Dibrugarh, Assam, 786004.\\$^2$Theoretical Physics Division, Centre for Atmospheric Studies, Dibrugarh University, Dibrugarh, Assam, 786004.\\}


\begin{abstract}
We present a dynamical systems approach to black hole thermodynamic criticality based on bifurcation equations.  We construct an effective thermodynamic landscape in which black holes relax toward equilibrium fixed points. To describe this process, we introduce a flow parameter $\tau$, interpreted as a phenomenological relaxation time, which governs the approach toward equilibrium configurations in thermodynamic state space.   Near critical points, the thermodynamic flow simplifies into universal mathematical forms, which allows different black holes to be grouped into different universality classes based on their critical behavior.  Our analysis further shows critical slowing down, with relaxation timescales determined entirely by the local bifurcation structure. 
\end{abstract}


\maketitle                                                                      

\section{Introduction}

The formulation of black hole thermodynamics \cite{Bekenstein:1973ur,Hawking:1974rv,Hawking:1975vcx,Bardeen:1973gs} has witnessed extensive developments and generalizations in different gravitational settings \cite{Wald:1979zz,bekenstein1980black,Wald:1999vt,Carlip:2014pma,Wall:2018ydq,Candelas:1977zz,Mahapatra:2011si}. Among the most actively studied topics are black hole phase transitions \cite{Davies:1989ey,Hawking:1982dh,curir_rotating_1981,Curir1981,Pavon:1988in,Pavon:1991kh,OKaburaki,Cai:1996df,Cai:1998ep,Wei:2009zzf,Bhattacharya:2019awq,Kastor:2009wy,Dolan:2010ha,Dolan:2011xt,Dolan:2011jm,Dolan:2012jh,Kubiznak:2012wp,Kubiznak:2016qmn}, including the Davies transition \cite{Davies:1989ey}, the Hawking--Page transition \cite{Hawking:1982dh}, extremal phase transitions \cite{curir_rotating_1981,Curir1981,Pavon:1988in,Pavon:1991kh,OKaburaki,Cai:1996df,Cai:1998ep,Wei:2009zzf,Bhattacharya:2019awq}, and Van der Waals type behavior in extended phase space \cite{Kastor:2009wy,Dolan:2010ha,Dolan:2011xt,Dolan:2011jm,Dolan:2012jh,Kubiznak:2012wp,Kubiznak:2016qmn}.  Recently, ideas from nonlinear dynamics have also found important applications in black hole physics. Dynamical stability during black hole phase transitions was investigated in Ref.~\cite{nl1}, while instabilities of thin black rings were studied in Ref.~\cite{nl2}. Chaotic behavior of geodesic motion induced by perturbations in black hole spacetimes was explored in Refs.~\cite{nl3,nl4}. Thermal chaos in black hole thermodynamics has further attracted considerable attention \cite{ch1,ch2,ch3,ch4}. In parallel, black hole phase transitions have also been analyzed from topological and phase-space viewpoints \cite{t1,t2,t3}.

Recent studies have explored the dynamical aspects of black hole phase transitions using stochastic evolution on the free energy landscape. In particular, Ref.~\cite{csd1} demonstrated that RN-AdS black holes exhibit pronounced critical slowing down near both critical and spinodal points, characterized by a sharp increase in the autocorrelation time and fluctuations of the order parameter due to the flattening of the free energy landscape. Extending this framework, Ref.~\cite{csd2} investigated AdS black holes and showed that the relaxation time near criticality follows a universal scaling relation, $\tau \sim |\epsilon|^{-2/3}$, suggesting the existence of universal dynamical behaviour across different black hole systems. These developments motivate the present work, where we further explore black hole critical dynamics from the perspective of bifurcation theory and thermodynamic relaxation behaviour.

\begin{figure}[t!]
\center
\includegraphics[scale=0.45]{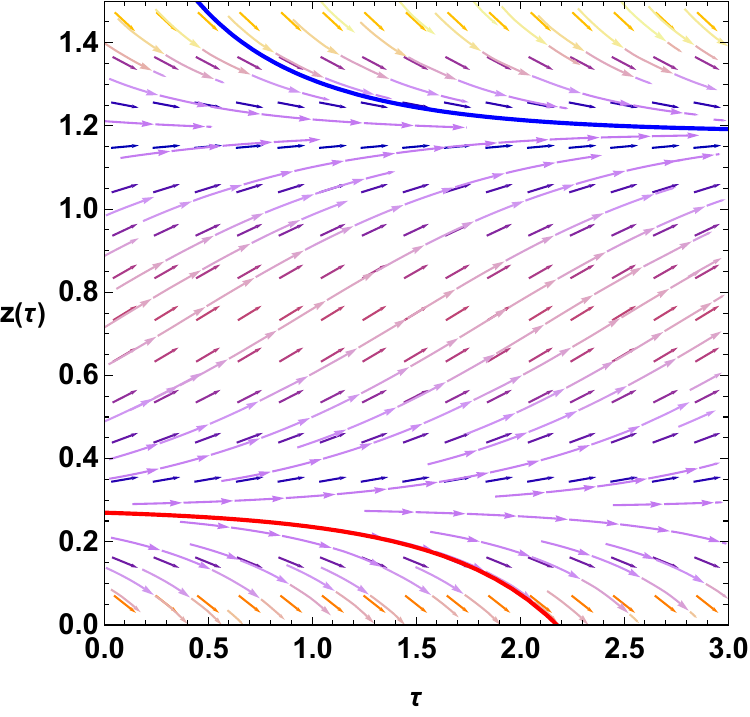}
\caption{Slowing down of black hole solution near fixed points}
\label{fig1}
\end{figure}
In one of our previous works \cite{bif}, we introduced a bifurcation-based perspective to study the emergence, stability, and criticality of black hole solutions. Building on the foundations of black hole thermodynamics, the work explored how the qualitative behavior of a black hole thermodynamic state changes as the system parameters are varied. In general, the solutions either converge to fixed points or diverge. However, the introduction of a control parameter gives rise to more interesting dynamical behavior. To facilitate this analysis, we introduce a \textit{bifurcation function} defined as  
\begin{equation}
\dot{z} = \frac{d z}{d \tau} = -\frac{d M}{d z} + h \frac{d S}{d z},
\label{master}
\end{equation}  
where \( h > 0 \) is a control parameter we refer to as the \textit{bifurcation parameter}, \( M \) denotes the mass, and \( S \) represents the black hole entropy. The variable \( z \) is a generalized measure of the black hole's size, which may correspond to quantities such as the event horizon radius or entropy. We also introduce an affine parameter \( \tau \), which serves to parameterize the evolution of the system and may be interpreted as an effective time variable in this dynamical framework.\\

While studying bifurcation and critical phenomena, we encountered some plots like Fig.~\ref{fig1}. These plots were obtained by numerically solving the bifurcation equation using initial conditions chosen from different thermodynamic branches of the black hole solution. For the stable branch, the solution gradually approaches the equilibrium configuration, whereas for the unstable branch, the trajectory diverges away from equilibrium. Another interesting feature observed in these plots is that the slope of the $z$--$\tau$ curve decreases as the solution approaches equilibrium. Since $z$ represents the size of the black hole and $\tau$ acts as an effective time parameter (not necessarily physical time), the slope may be interpreted as a velocity-like quantity associated with the thermodynamic transition process. Consequently, as the black hole configuration approaches the critical point or fixed point, the evolution naturally exhibits a slowing-down behavior.
This observation motivates the following question \textbf{Can the rate at which this slowing down occurs be classified into universal classes?}

Guided by this question, we investigate the critical slowing-down behavior associated with thermodynamic bifurcations in black holes.

 \section*{Derivation of Critical Slowing Down from the Bifurcation Function}
 \begin{table*}[t]
\centering
\setlength{\tabcolsep}{6pt}
\begin{tabular}{|c|c|c|c|c|}
\hline
\textbf{Black Hole System} & \textbf{No. of Fixed Points} & \textbf{No. of Half-Stable Points} & \textbf{Bifurcation Type} \\
\hline
Schwarzschild-AdS & 2 & 1 & Saddle-node \\
\hline
RN AdS & 3 or 1 & 2 & Broken pitchfork \\
\hline
Euler Heisenberg AdS & 4, 2 or 0 & 3 & 3-Multifold saddle node \\
\hline
6D Gauss Bonnet AdS & 5, 3 or 1 & 4 & 4-Multifold saddle-node \\
\hline
\end{tabular}
\caption{Summary of bifurcation structures in various black hole systems studied in \cite{bif}.}
\label{tab1}
\end{table*}

Consider a general form of the bifurcation equation \ref{master} as 
\begin{equation}
\dot{z}=F(z,h,x_i),
\end{equation}
where $x_i$ denotes all other black-hole-specific parameters (charge/coupling constants, etc.).\\

The equilibrium configurations are determined from
\begin{equation}
F(z,h,x_i)=0.
\end{equation}

At the critical point $(z_c,h_c)$, the bifurcation conditions are
\begin{equation}
F(z_c,h_c,x_i)=0,
\end{equation}
and
\begin{equation}
\left.\frac{\partial F}{\partial z}\right|_{(z_c,h_c)}=0.
\end{equation}

The second condition implies that the linear restoring force vanishes at criticality.

Let
\begin{equation}
z=z_*+\delta,
\end{equation}
where $z_*$ is an equilibrium point satisfying
\begin{equation}
F(z_*,h,x_i)=0.
\end{equation}

Substituting into the dynamical equation,
\begin{equation}
\frac{d \delta}{d \tau}=\dot{\delta}=F(z_*+\delta,h,x_i).
\end{equation}

Expanding to linear order,
\begin{equation}
F(z_*+\delta,h,x_i)
=
F(z_*,h,x_i)
+
\left.
\frac{\partial F}{\partial z}
\right|_{z_*}\delta
+\mathcal{O}(\delta^2).
\end{equation}

Using $F(z_*,h,x_i)=0$, one obtains
\begin{equation}
\dot{\delta}=\lambda \delta,
\end{equation}
where
\begin{equation}
\lambda=
\left.
\frac{\partial F}{\partial z}
\right|_{z_*}.
\end{equation}

The solution is
\begin{equation}
\delta(t)=\delta_0 e^{\lambda \tau}.
\end{equation}

Thus the characteristic relaxation timescale is
\begin{equation}
\tau=\frac{1}{|\lambda|}.
\end{equation}

As the critical point is approached,
\begin{equation}
\lambda\rightarrow0,
\end{equation}
which implies
\begin{equation}
\tau\rightarrow\infty.
\end{equation}

This divergence of the relaxation time is the phenomenon of critical slowing down.\\

Next, we Introduce deviations from the critical point:
\begin{equation}
z=z_c+\xi,
\qquad
h=h_c+\mu,
\end{equation}
where
\begin{equation}
\mu=h-h_c
\end{equation}
measures the distance from criticality.

Expanding the bifurcation function near the critical point,
\begin{equation}
F(z_c+\xi,h_c+\mu,x_i)
=
F_h \mu
+
\frac{1}{n!}F^{(n)}\xi^n
+\cdots,
\end{equation}
where $n$ is the first non-vanishing derivative order satisfying
\begin{equation}
F_z=F_{zz}=\cdots=F^{(n-1)}=0,
\qquad
F^{(n)}\neq0.
\end{equation}

After rescaling variables, the universal normal form becomes
\begin{equation}
\dot{\xi}=\mu+\xi^n.
\end{equation}

The equilibrium condition is
\begin{equation}
0=\mu+\xi^n,
\end{equation}
which gives
\begin{equation}
\xi_*\sim \mu^{1/n}.
\end{equation}

The corresponding stability eigenvalue is
\begin{equation}
\lambda=
\frac{\partial \dot{\xi}}{\partial \xi}
=n\xi^{n-1}.
\end{equation}

Using the equilibrium scaling,
\begin{equation}
\lambda\sim \mu^{(n-1)/n}.
\end{equation}

Finally, the relaxation time scales as
\begin{equation}
\tau\sim \mu^{-(n-1)/n}.
\end{equation}

Thus different orders of degeneracy produce distinct dynamical universality classe

In the next section,  we investigate the critical slowing-down behavior associated with thermodynamic bifurcations in black holes. To illustrate the generality of the framework, we consider four representative systems: Schwarzschild-AdS, RN-AdS, Euler--Heisenberg-AdS, and 6D Gauss-Bonnet-AdS black holes. These examples are specifically chosen because they exhibit different bifurcation structures and distinct numbers of fixed points and half-stable points, as summarized in Table~\ref{tab1}. Correspondingly, the associated thermodynamic flows belong to different universality classes, ranging from the standard saddle-node bifurcation to higher-order multifold saddle-node structures. We show that these differences naturally lead to distinct critical slowing-down exponents governed entirely by the local bifurcation structure near the critical point.

\section{Schwarzschild-AdS Black Hole and Critical Slowing Down}

We begin with the bifurcation function
\begin{equation}
\dot{z}=F(z,h),
\end{equation}
where $F(z,h)$ for the Schwarzschild-AdS black hole is
\begin{equation}
F(z,h)=\frac{1}{2}\left(1-4\pi h z+3z^2\right).
\end{equation}

The equilibrium points of the system are obtained from
\begin{equation}
F(z,h)=0.
\end{equation}

The critical point occurs when two equilibrium branches merge together. Mathematically, this happens when
\begin{equation}
F(z_c,h_c)=0,
\qquad
\frac{\partial F}{\partial z}\Big|_{(z_c,h_c)}=0.
\end{equation}

The derivative of the bifurcation function is
\begin{equation}
\frac{\partial F}{\partial z}=-2\pi h+3z.
\end{equation}

Setting this equal to zero gives
\begin{equation}
z_c=\frac{2\pi h_c}{3}.
\end{equation}

Substituting this into the equilibrium equation, we obtain
\begin{equation}
1-\frac{4\pi^2 h_c^2}{3}=0.
\end{equation}

Hence the critical point is
\begin{equation}
h_c=\frac{\sqrt{3}}{2\pi},
\qquad
z_c=\frac{1}{\sqrt{3}}.
\end{equation}

To study the behavior near the critical point, we introduce small deviations
\begin{equation}
z=z_c+\xi,
\qquad
h=h_c+\mu,
\end{equation}
where
\begin{equation}
\mu=h-h_c
\end{equation}
measures the distance from the critical point.

Substituting these into the bifurcation equation and expanding near the critical point, the leading-order dynamics becomes
\begin{equation}
\dot{\xi}
=
F_h|_c\,\mu
+
\frac{1}{2}F_{zz}|_c\,\xi^2.
\end{equation}

The required derivatives are
\begin{equation}
F_h=-2\pi z,
\qquad
F_{zz}=3.
\end{equation}

Evaluating them at the critical point,
\begin{equation}
F_h|_c=-\frac{2\pi}{\sqrt{3}},
\qquad
F_{zz}|_c=3.
\end{equation}

Therefore,
\begin{equation}
\dot{\xi}
=
-\frac{2\pi}{\sqrt{3}}\mu
+
\frac{3}{2}\xi^2.
\end{equation}

After rescaling the variables, this equation takes the universal form
\begin{equation}
\dot{\xi}=\mu+\xi^2.
\label{normal}
\end{equation}

This is the standard saddle-node bifurcation normal form.

The equilibrium solutions are obtained from
\begin{equation}
\mu+\xi^2=0.
\end{equation}

Thus,
\begin{equation}
\xi_*\sim \mu^{1/2}.
\end{equation}

Next, we perturb the equilibrium solution by writing
\begin{equation}
\xi=\xi_*+\delta.
\end{equation}

Substituting into the bifurcation equation and keeping only linear terms in $\delta$, we obtain
\begin{equation}
\dot{\delta}=\lambda\delta,
\end{equation}
where
\begin{equation}
\lambda=
\frac{\partial \dot{\xi}}{\partial \xi}\Big|_{\xi_*}.
\end{equation}

Using the normal form equation \ref{normal},
\begin{equation}
\dot{\xi}=\mu+\xi^2,
\end{equation}
gives
\begin{equation}
\lambda=2\xi_*.
\end{equation}

Since
\begin{equation}
\xi_*\sim\mu^{1/2},
\end{equation}
the eigenvalue behaves as
\begin{equation}
\lambda\sim\mu^{1/2}.
\end{equation}

The relaxation time is defined as
\begin{equation}
\tau=\frac{1}{|\lambda|}.
\end{equation}

Therefore,
\begin{equation}
\tau\sim\mu^{-1/2}.
\end{equation}

Hence, as the critical point is approached $(\mu\to0)$, the relaxation time diverges. This is the phenomenon of critical slowing down.

\section{RN-AdS Black Hole and Higher-Order Criticality}

For the RN-AdS black hole, the thermodynamic flow equation is
\begin{equation}
\dot{z}=F(z,h),
\end{equation}
with bifurcation function
\begin{equation}
F(z,h)=\frac{1}{2}\left(-1+\frac{Q^2}{z^2}+4\pi h z-3z^2\right).
\end{equation}

The equilibrium configurations are obtained from
\begin{equation}
F(z,h)=0.
\end{equation}

The critical point is determined from the degeneracy conditions
\begin{equation}
F(z_c,h_c)=0,
\qquad
F_z(z_c,h_c)=0,
\qquad
F_{zz}(z_c,h_c)=0.
\end{equation}

To study the dynamics near criticality, we introduce
\begin{equation}
z=z_c+\zeta,
\qquad
h=h_c+\mu,
\end{equation}
where $\mu$ measures the distance from the critical point.

Expanding the flow equation near the critical point, the linear and quadratic terms vanish due to the degeneracy conditions. The leading-order dynamics therefore becomes
\begin{equation}
\dot{\zeta}=a\mu+b \zeta^3.
\end{equation}

After suitable rescaling, this reduces to the universal cusp normal form
\begin{equation}
\dot{\zeta}=\mu+\zeta^3.
\label{normal2}
\end{equation}

The equilibrium solutions satisfy
\begin{equation}
\mu+\zeta_*^3=0,
\end{equation}
which gives
\begin{equation}
\zeta_*\sim \mu^{1/3}.
\end{equation}

Now perturb the equilibrium solution as
\begin{equation}
\zeta=\zeta_*+\delta.
\end{equation}

Substituting into  equation \ref{normal2} and keeping only linear terms in $\delta$, we obtain
\begin{equation}
\lambda=3\zeta_*^2.
\end{equation}

Since
\begin{equation}
zeta_*\sim\mu^{1/3},
\end{equation}
the eigenvalue scales as
\begin{equation}
\lambda\sim\mu^{2/3}.
\end{equation}

Therefore the relaxation timescale behaves as
\begin{equation}
\tau_{\rm relax}\sim\mu^{-2/3}.
\end{equation}

Thus, unlike the Schwarzschild-AdS black hole which belongs to the saddle-node bifurcations, the RN-AdS black hole exhibits broken pitchfork bifurcation characterized by a cubic normal form and a different slowing-down exponent.
\section{Euler-Heisenberg Black Hole and Higher-Order Criticality}

For the Euler-Heisenberg black hole bifurcation function is
\begin{equation}
F(z,h)=\frac{1}{8}\left[4\left(-1+\frac{q^2}{z^2}+4\pi h z-3z^2\right)-\frac{\tilde{\alpha} q^4}{z^6}\right].
\end{equation}

The equilibrium configurations are determined from
\begin{equation}
F(z,h)=0.
\end{equation}

The critical point is obtained from the higher degeneracy conditions
\begin{eqnarray}
F(z_c,h_c)=0,
\qquad
F_z(z_c,h_c)=0,\\ \nonumber
F_{zz}(z_c,h_c)=0,
\qquad
F_{zzz}(z_c,h_c)=0.
\end{eqnarray}

To study the dynamics near criticality, we introduce
\begin{equation}
z=z_c+\zeta,
\qquad
h=h_c+\mu,
\end{equation}
where $\mu$ measures the distance from the critical point.

Expanding the flow equation near the critical point, the linear, quadratic and cubic terms vanish due to the degeneracy conditions. The leading-order dynamics therefore becomes
\begin{equation}
\dot{\zeta}=a\mu+b\zeta^4.
\end{equation}

After suitable rescaling, this reduces to the universal quartic normal form
\begin{equation}
\dot{\zeta}=\mu+\zeta^4.
\label{normal3}
\end{equation}

The equilibrium solutions satisfy
\begin{equation}
\mu+\zeta_*^4=0,
\end{equation}
which gives
\begin{equation}
\zeta_*\sim\mu^{1/4}.
\end{equation}

Now perturb the equilibrium solution as
\begin{equation}
\zeta=\zeta_*+\delta.
\end{equation}

Substituting into Eq.~(\ref{normal3}) and keeping only linear terms in $\delta$, we obtain
\begin{equation}
\lambda=4\zeta_*^3.
\end{equation}

Since
\begin{equation}
\zeta_*\sim\mu^{1/4},
\end{equation}
the eigenvalue scales as
\begin{equation}
\lambda\sim\mu^{3/4}.
\end{equation}

Therefore the relaxation timescale behaves as
\begin{equation}
\tau_{\rm relax}\sim\mu^{-3/4}.
\end{equation}

Thus the Euler-Heisenberg black hole exhibits a higher-order thermodynamic bifurcation characterized by a quartic normal form and a distinct critical slowing-down exponent.\\
\section{6D Gauss-Bonnet Black Hole}

For the 6D Gauss-Bonnet black hole, the thermodynamic flow equation is
\begin{equation}
\dot{r}=F(r,h),
\end{equation}
with bifurcation function
\begin{equation}
F(r,h)=
-\frac{-3 q^2 + 18 r^6 - 32 \pi h r^7 + 30 r^8 + 36 r^4 \tilde{\alpha}
-384 \pi h r^5 \tilde{\alpha}}
{32 \pi r^4}.
\end{equation}

The critical point satisfies
\begin{eqnarray}
F(r_c,h_c)=0,
\qquad
F_r(r_c,h_c)=0,\\ \nonumber
F_{rr}(r_c,h_c)=0,
\qquad
F_{rrr}(r_c,h_c)=0.
\end{eqnarray}

Expanding near the critical point,
\begin{equation}
r=r_c+\zeta,
\qquad
h=h_c+\mu,
\end{equation}
the leading-order dynamics becomes
\begin{equation}
\dot{\zeta}=\mu+\zeta^4.
\end{equation}

Therefore,
\begin{equation}
\zeta_*\sim\mu^{1/4},
\qquad
\lambda\sim\mu^{3/4},
\qquad
\tau_{\rm relax}\sim\mu^{-3/4}.
\end{equation}

Thus the 6D Gauss-Bonnet black hole belongs to the same quartic universality class as the Euler-Heisenberg black hole.

\section{Conclusion}

In this work, we investigated black hole thermodynamic criticality from the perspective of nonlinear dynamical systems and bifurcation theory. By constructing a effective bifurcation equations, we interpreted equilibrium black hole configurations as fixed points of a thermodynamic landscape and introduced a phenomenological flow parameter that governs the relaxation toward equilibrium states.

We showed that different black hole systems shows distinct bifurcations characterized by different orders of degeneracy. The Schwarzschild-AdS black hole displays the standard saddle-node bifurcation with relaxation scaling
\begin{equation}
\tau_{\rm relax}\sim \mu^{-1/2},
\end{equation}
while the RN-AdS black hole exhibits a broken pitchfork-type structure with
\begin{equation}
\tau_{\rm relax}\sim \mu^{-2/3}.
\end{equation}
For Euler--Heisenberg-AdS and 6D Gauss--Bonnet-AdS black holes, higher-order multifold saddle-node structures emerge, leading to
\begin{equation}
\tau_{\rm relax}\sim \mu^{-3/4}.
\end{equation}
A larger magnitude of the exponent implies a stronger delay in the relaxation process near criticality. Since the relaxation time diverges faster for larger exponents, higher-order bifurcation structures correspond to increasingly delayed approaches toward equilibrium near the critical point. In this sense, the slowing-down exponent quantifies the degree of critical bottlenecking induced by the local degeneracy of the thermodynamic landscape.

Our results demonstrated that the critical slowing down exponent can be determined entirely by the local bifurcation structure near the critical point.  Consequently, distinct black hole systems may belong to the same dynamical universality class despite possessing different global phase structures.
The present framework provides a simple dynamical interpretation of black hole phase transitions,  we hope that this approach may offer further insights into universal aspects of gravitational thermodynamics and related dynamical processes.


\begin{thebibliography}{99}
		
		
	\def\EPJC{Eur. Phys. J. C\,}
\def\IJMPA{Int. J. Mod. Phys. A\,}
\def\JCAP{J. Cosmol. Astropart. Phys.\,}
\def\JHEP{J. High Energy Phys.\,}
\def\CQG{Classical Quantum Gravity\,}
\def\JMP{J. Math. Phys. (N.Y.)\,}
\def\NPB{Nucl. Phys. B \,}
\def\PDU{Phys. Dark Univ.\,}
\def\PLB{Phys. Lett. B \,}
\def\PRD{Phys. Rev. D\,}
\def\PRL{Phys. Rev. Lett.\,}
\def\PRR{Phys. Rev. Res.\,}
\def\GRG{Gen. Relativ. Gravit.\,}
	
	
		\bibitem{Bekenstein:1973ur}
		J.~D.~Bekenstein,
		Black holes and entropy,
		Phys. Rev. D \textbf{7}, 2333-2346 (1973)
		doi:10.1103/PhysRevD.7.2333

		\bibitem{Hawking:1974rv}
		S.~W.~Hawking,
		Black hole explosions,
		Nature \textbf{248}, 30-31 (1974)
		doi:10.1038/248030a0
		\bibitem{Hawking:1975vcx}
		S.~W.~Hawking,
		Particle Creation by Black Holes,
		Commun. Math. Phys. \textbf{43}, 199-220 (1975)
		[erratum: Commun. Math. Phys. \textbf{46}, 206 (1976)]
		doi:10.1007/BF02345020
		
		
		
		
		
		
		
		
		\bibitem{Bardeen:1973gs}
		J.~M.~Bardeen, B.~Carter and S.~W.~Hawking,
		The Four laws of black hole mechanics,
		Commun. Math. Phys. \textbf{31}, 161-170 (1973)
		doi:10.1007/BF01645742
		
		
		
		
		
		
		\bibitem{Wald:1979zz}
		R.~M.~Wald,
		Entropy and black-hole thermodynamics,
		Phys. Rev. D \textbf{20}, 1271-1282 (1979)
		doi:10.1103/PhysRevD.20.1271
		\bibitem{bekenstein1980black}
		Jacob~D Bekenstein.
		Black-hole thermodynamics,
		Physics Today, 33(1):24--31, 1980.
		\bibitem{Wald:1999vt}
		R.~M.~Wald,
		The thermodynamics of black holes,
		Living Rev. Rel. \textbf{4}, 6 (2001)
		doi:10.12942/lrr-2001-6
		[arXiv:gr-qc/9912119 [gr-qc]].
		
		\bibitem{Carlip:2014pma}
		S.Carlip,
		Black Hole Thermodynamics,
		Int. J. Mod. Phys. D \textbf{23}, 1430023 (2014)
		doi:10.1142/S0218271814300237
		[arXiv:1410.1486 [gr-qc]].
		\bibitem{Wall:2018ydq}
		A.C.Wall,
		A Survey of Black Hole Thermodynamics,
		[arXiv:1804.10610 [gr-qc]].
		\bibitem{Candelas:1977zz}
		P.Candelas and D.W.Sciama,
		Irreversible Thermodynamics of Black Holes,
		Phys. Rev. Lett. \textbf{38}, 1372-1375 (1977)
		doi:10.1103/PhysRevLett.38.1372
		\bibitem{Mahapatra:2011si}
		S.~Mahapatra, P.~Phukon and T.~Sarkar,
		Phys. Rev. D \textbf{84}, 044041 (2011)
		doi:10.1103/PhysRevD.84.044041
		[arXiv:1103.5885 [hep-th]].
	
	
	
	
	
	
		\bibitem{Davies:1989ey}
		P.~C.~W.~Davies,
		Thermodynamic Phase Transitions of {Kerr-Newman} Black Holes in De Sitter Space,
		Class. Quant. Grav. \textbf{6}, 1909 (1989)
		doi:10.1088/0264-9381/6/12/018
		
		
		
		
		
		
		
		
		\bibitem{Hawking:1982dh}
		S.~W.~Hawking and D.~N.~Page,
		Thermodynamics of Black Holes in anti-De Sitter Space,
		Commun. Math. Phys. \textbf{87}, 577 (1983)
		doi:10.1007/BF01208266
		
		
		
		
		
		
		
		\bibitem{curir_rotating_1981}
		A. Curir,
		Rotating black holes as dissipative spin-thermodynamical systems,
		General Relativity and Gravitation,
		\textbf{13}, 417, (1981)
		doi:10.1007/BF00756588
		\bibitem{Curir1981}
		Anna Curir, Black hole emissions and phase transitions,
		General Relativity and Gravitation,
		\textbf{13}, 1177, (1981)
		doi:10.1007/BF00759866
		\bibitem{Pavon:1988in}
		D.~Pavon and J.~M.~Rubi,
		Nonequilibrium Thermodynamic Fluctuations of Black Holes,
		Phys. Rev. D \textbf{37}, 2052-2058 (1988)
		doi:10.1103/PhysRevD.37.2052
		\bibitem{Pavon:1991kh}
		D.~Pavon,
		Phase transition in Reissner-Nordstrom black holes,
		Phys. Rev. D \textbf{43}, 2495-2497 (1991)
		doi:10.1103/PhysRevD.43.2495
		\bibitem{OKaburaki}
		O.~Kaburaki, Critical behavior of extremal Kerr-Newman black holes,
		Gen. Rel. Grav. \textbf{28}, 843 (1996)
		\bibitem{Cai:1996df}
		R.~G.~Cai, Z.~J.~Lu and Y.~Z.~Zhang,
		Critical behavior in (2+1)-dimensional black holes,
		Phys. Rev. D \textbf{55}, 853-860 (1997)
		doi:10.1103/PhysRevD.55.853
		[arXiv:gr-qc/9702032 [gr-qc]].
		\bibitem{Cai:1998ep}
		R.~G.~Cai and J.~H.~Cho,
		Thermodynamic curvature of the BTZ black hole,
		Phys. Rev. D \textbf{60}, 067502 (1999)
		doi:10.1103/PhysRevD.60.067502
		[arXiv:hep-th/9803261 [hep-th]].
		\bibitem{Wei:2009zzf}
		Y.~H.~Wei,
		Thermodynamic critical and geometrical properties of charged BTZ black hole,
		Phys. Rev. D \textbf{80}, 024029 (2009)
		doi:10.1103/PhysRevD.80.024029
		\bibitem{Bhattacharya:2019awq}
		K.~Bhattacharya, S.~Dey, B.~R.~Majhi and S.~Samanta,
		General framework to study the extremal phase transition of black holes,
		Phys. Rev. D \textbf{99}, no.12, 124047 (2019)
		doi:10.1103/PhysRevD.99.124047
		[arXiv:1903.03434 [gr-qc]].
		
		
		
		
		
		
		
		
		\bibitem{Kastor:2009wy}
		D.~Kastor, S.~Ray and J.~Traschen,
		Enthalpy and the Mechanics of AdS Black Holes,
		Class. Quant. Grav. \textbf{26}, 195011 (2009)
		doi:10.1088/0264-9381/26/19/195011
		[arXiv:0904.2765 [hep-th]].
		\bibitem{Dolan:2010ha}
		B.~P.~Dolan,
		The cosmological constant and the black hole equation of state,
		Class. Quant. Grav. \textbf{28}, 125020 (2011)
		doi:10.1088/0264-9381/28/12/125020
		[arXiv:1008.5023 [gr-qc]].
		\bibitem{Dolan:2011xt}
		B.~P.~Dolan,
		Pressure and volume in the first law of black hole thermodynamics,
		Class. Quant. Grav. \textbf{28}, 235017 (2011)
		doi:10.1088/0264-9381/28/23/235017
		[arXiv:1106.6260 [gr-qc]].
		\bibitem{Dolan:2011jm}
		B.~P.~Dolan,
		Compressibility of rotating black holes,
		Phys. Rev. D \textbf{84}, 127503 (2011)
		doi:10.1103/PhysRevD.84.127503
		[arXiv:1109.0198 [gr-qc]].
		\bibitem{Dolan:2012jh}
		B.~P.~Dolan,
		Where Is the PdV in the First Law of Black Hole Thermodynamics?,
		doi:10.5772/52455
		[arXiv:1209.1272 [gr-qc]].
		\bibitem{Kubiznak:2012wp}
		D.~Kubiznak and R.~B.~Mann,
		P-V criticality of charged AdS black holes,
		JHEP \textbf{07}, 033 (2012)
		doi:10.1007/JHEP07(2012)033
		[arXiv:1205.0559 [hep-th]].
		\bibitem{Kubiznak:2016qmn}
		D.~Kubiznak, R.~B.~Mann and M.~Teo,
		Black hole chemistry: thermodynamics with Lambda,
		Class. Quant. Grav. \textbf{34}, no.6, 063001 (2017)
		doi:10.1088/1361-6382/aa5c69
		[arXiv:1608.06147 [hep-th]].
			\bibitem{rp1} J.~Sadeghi, M.~Shokri, S.~Gashti Noori and M.~R.~Alipour,
RPS thermodynamics of Taub\textendash{}NUT AdS black holes in the presence of central charge and the weak gravity conjecture.
		Gen. Rel. Grav. \textbf{54} (2022)
		\bibitem{rp2}
		Y.~Ladghami, B.~Asfour, A.~Bouali, A.~Errahmani and T.~Ouali,
4D-EGB black holes in RPS thermodynamics,
		Phys. Dark Univ. \textbf{41} (2023),
		\bibitem{rp3}
		X.~Kong, Z.~Zhang and L.~Zhao,
		Restricted phase space thermodynamics of charged AdS black holes in conformal gravity,
		Chin. Phys. C \textbf{47} (2023)
		\bibitem{rp4}
		M.~R.~Alipour, J.~Sadeghi and M.~Shokri,
		WGC and WCCC of black holes with quintessence and cloud strings in RPS space,
		Nucl. Phys. B \textbf{990} (2023)
		\bibitem{rp5}
		T.~Wang and L.~Zhao,
		Black hole thermodynamics is extensive with variable Newton constant,
		Phys. Lett. B \textbf{827} (2022)
		\bibitem{rp6}
		G.~Zeyuan and L.~Zhao,
		Restricted phase space thermodynamics for AdS black holes via holography,
		Class. Quant. Grav. \textbf{39} (2022)
		\bibitem{rp7}
		Z.~Gao, X.~Kong and L.~Zhao,
	Thermodynamics of Kerr-AdS black holes in the restricted phase space,
		Eur. Phys. J. C \textbf{82} (2022)
		\bibitem{rp8}
		S.~Dutta and G.~S.~Punia,
		String theory corrections to holographic black hole chemistry,
		Phys. Rev. D \textbf{106}, no.2, 026003 (2022)
		\bibitem{rp9}
		T.~F.~Gong, J.~Jiang and M.~Zhang,
		Holographic thermodynamics of rotating black holes,
		JHEP \textbf{06}, 105 (2023)
		\bibitem{rp10}
		W.~Cong, D.~Kubiznak, R.~B.~Mann and M.~R.~Visser,
		Holographic CFT phase transitions and criticality for charged AdS black holes,
		JHEP \textbf{08}, 174 (2022)
		\bibitem{rp11}
		M.~R.~Visser,
		Holographic thermodynamics requires a chemical potential for color,
		Phys. Rev. D \textbf{105}, no.10, 106014 (2022)
		
		
		
		
		
		\bibitem{Quevedo:2008ry}
		H.~Quevedo and A.~Sanchez,
		Geometric description of BTZ black holes thermodynamics,
		Phys. Rev. D \textbf{79}, 024012 (2009)
		doi:10.1103/PhysRevD.79.024012
		[arXiv:0811.2524 [gr-qc]].
		\bibitem{Akbar:2011qw}
		M.~Akbar, H.~Quevedo, K.~Saifullah, A.~Sanchez and S.~Taj,
		Thermodynamic Geometry Of Charged Rotating BTZ Black Holes,
		Phys. Rev. D \textbf{83}, 084031 (2011)
		doi:10.1103/PhysRevD.83.084031
		[arXiv:1101.2722 [gr-qc]].
		\bibitem{Hendi:2015cka}
		S.~H.~Hendi and R.~Naderi,
		Geometrothermodynamics of black holes in Lovelock gravity with a nonlinear electrodynamics,
		Phys. Rev. D \textbf{91}, no.2, 024007 (2015)
		doi:10.1103/PhysRevD.91.024007
		[arXiv:1510.06269 [hep-th]].
		\bibitem{Sarkar:2006tg}
		T.~Sarkar, G.~Sengupta and B.~Nath Tiwari,
		On the thermodynamic geometry of BTZ black holes,
		JHEP \textbf{11}, 015 (2006)
		doi:10.1088/1126-6708/2006/11/015
		[arXiv:hep-th/0606084 [hep-th]].
		\bibitem{Hendi:2015xya}
		S.~H.~Hendi, A.~Sheykhi, S.~Panahiyan and B.~Eslam Panah,
		Phase transition and thermodynamic geometry of Einstein-Maxwell-dilaton black holes,
		Phys. Rev. D \textbf{92}, no.6, 064028 (2015)
		doi:10.1103/PhysRevD.92.064028
		[arXiv:1509.08593 [hep-th]].
		\bibitem{Banerjee:2016nse}
		R.~Banerjee, B.~R.~Majhi and S.~Samanta,
		Thermogeometric phase transition in a unified framework,
		Phys. Lett. B \textbf{767}, 25-28 (2017)
		doi:10.1016/j.physletb.2017.01.040
		[arXiv:1611.06701 [gr-qc]].
		\bibitem{Bhattacharya:2017hfj}
		K.~Bhattacharya and B.~R.~Majhi,
		Thermogeometric description of the van der Waals like phase transition in AdS black holes,
		Phys. Rev. D \textbf{95}, no.10, 104024 (2017)
		doi:10.1103/PhysRevD.95.104024
		[arXiv:1702.07174 [gr-qc]].
		\bibitem{Bhattacharya:2019qxe}
		K.~Bhattacharya and B.~R.~Majhi,
		Thermogeometric study of van der Waals like phase transition in black holes: an alternative approach,
		Phys. Lett. B \textbf{802}, 135224 (2020)
		doi:10.1016/j.physletb.2020.135224
		[arXiv:1903.10370 [gr-qc]].
		
		\bibitem{Gogoi:2023qni}
		N.~J.~Gogoi, G.~K.~Mahanta and P.~Phukon,
		Geodesics in geometrothermodynamics (GTD) type II geometry of 4D asymptotically anti-de-Sitter black holes,
		Eur. Phys. J. Plus \textbf{138}, no.4, 345 (2023)
		doi:10.1140/epjp/s13360-023-03938-x
		
		\bibitem{Kumar:2012ve}
		P.~Kumar, S.~Mahapatra, P.~Phukon and T.~Sarkar,
		Geodesics in Information Geometry : Classical and Quantum Phase Transitions,
		Phys. Rev. E \textbf{86}, 051117 (2012)
		doi:10.1103/PhysRevE.86.051117
		[arXiv:1210.7135 [cond-mat.stat-mech]].
		
\bibitem{Wei:2015iwa}
S.~W.~Wei and Y.~X.~Liu,
Insight into the microscopic structure of an AdS black hole from a thermodynamical phase transition,
\emph{Phys. Rev. Lett.} \textbf{115}, 111302 (2015),
doi:10.1103/PhysRevLett.115.111302
[arXiv:1502.00386 [gr-qc]].

\bibitem{Altamirano:2013uqa}
N.~Altamirano, D.~Kubiznak, R.~B.~Mann and Z.~Sherkatghanad,
Kerr-AdS analogue of triple point and solid/liquid/gas phase transition,
\emph{Class. Quant. Grav.} \textbf{31}, 042001 (2014),
doi:10.1088/0264-9381/31/4/042001
[arXiv:1308.2672 [hep-th]].

\bibitem{Hennigar:2017apu}
R.~A.~Hennigar, R.~B.~Mann and E.~Tjoa,
Superfluid Black Holes,
\emph{Phys. Rev. Lett.} \textbf{118}, 021301 (2017),
doi:10.1103/PhysRevLett.118.021301
[arXiv:1609.02564 [hep-th]].

\bibitem{nl1}
G.~Arcioni and E.~Lozano-Tellechea, Stability and critical phenomena of black holes and black rings,
Phys. Rev. D \textbf{72}, 104021 (2005)
\bibitem{nl2}
J.~Armas and E.~Parisini,Instabilities of Thin Black Rings: Closing the Gap,''
JHEP \textbf{04}, 169 (2019)
\bibitem{nl3}
L.~Polcar, P.~Sukov{\'a} and O.~Semer{\'a}k, Free Motion around Black Holes with Disks or Rings: Between Integrability and Chaos{\textendash}V,'
Astrophys. J. \textbf{877}, no.1, 16 (2019)
\bibitem{nl4}
T.~S.~Amancio, R.~A.~Mosna and R.~S.~S.~Vieira, Chaotic orbital dynamics of pulsating stars around black holes surrounded by dark matter halos,Phys. Rev. D \textbf{110}, no.12, 124048 (2024)




\bibitem{ch1} Yong Chen, Haitang Li, and Shao-Jun Zhang. Chaos in Born–Infeld–AdS black hole
within extended phase space. Gen. Rel. Grav., 51(10):134, 2019.
\bibitem{ch2} Sandip Mahish and Bhamidipati Chandrasekhar. Chaos in Charged Gauss-Bonnet AdS
Black Holes in Extended Phase Space. Phys. Rev. D, 99(10):106012, 2019.
\bibitem{ch3} Chaoqun Dai, Songbai Chen, and Jiliang Jing. Thermal chaos of a charged dilaton-AdS
black hole in the extended phase space. Eur. Phys. J. C, 80(3):245, 2020.
\bibitem{ch4} Bing Tang. Temporal and spatial chaos in the Kerr-AdS black hole in an extended
phase space. Chin. Phys. C, 45(5):055101, 2021.

\bibitem{t1}
	S.~W.~Wei and Y.~X.~Liu,
	Topology of black hole thermodynamics,
	Phys. Rev. D \textbf{105}, no.10, 104003 (2022)
\bibitem{t2}
S.~W.~Wei, Y.~X.~Liu and R.~B.~Mann, Black Hole Solutions as Topological Thermodynamic Defects,
Phys. Rev. Lett. \textbf{129}, no.19, 191101 (2022)
\bibitem{t3}
D.~Wu, W.~Liu, S.~Q.~Wu and R.~B.~Mann, Novel topological classes in black hole thermodynamics,
Phys. Rev. D \textbf{111}, no.6, L061501 (2025)
\bibitem{csd1}  Li, Ran and Zhang, Kun and Yang, Jiayue and Mann, Robert B. and Wang, Jin, Critical slowing down of black hole phase transition and kinetic crossover in supercritical regime,  Phys. Rev. D, \textbf{112},6, 064004, 2025.
\bibitem{csd2} Mozib Bin Awal,  Prabwal Phukon,  Critical slowing down of black hole phase transition and universal dynamic scaling in AdS black holes, 	arXiv:2605.15655 [hep-th]
\bibitem{bif} B.  Hazarika,  P.  Phukon,  Bifurcation and critical phenomena in black hole thermodynamics. Eur. Phys. J. C 85, 1015 (2025). https://doi.org/10.1140/epjc/s10052-025-14744-3
 
\end{thebibliography}
\end{document}